\documentclass{article}
\usepackage{ijcai26}

\usepackage{times}
\usepackage{soul}
\usepackage{url}
\usepackage[hidelinks]{hyperref}
\usepackage[utf8]{inputenc}
\usepackage[small]{caption}
\usepackage{graphicx}
\usepackage{amsmath}
\usepackage{amsthm}
\usepackage{booktabs}
\usepackage{algorithm}
\usepackage{algorithmic}
\usepackage{multirow}
\usepackage{pifont}
\usepackage{xspace}
\usepackage{amssymb}

\makeatletter
\DeclareRobustCommand\onedot{\futurelet\@let@token\@onedot}
\def\@onedot{\ifx\@let@token.\else.\null\fi\xspace}
\def\eg{\emph{e.g}\onedot} 
\def\ie{\emph{i.e}\onedot} 
 
\def\etc{\emph{etc}\onedot} 
\def\etal{\emph{et al}\onedot}

\makeatother

\newcommand{\Name}{VeilNN\xspace}
\def\eg{\emph{e.g}\onedot} 
\RequirePackage[font=footnotesize,skip=3pt,subrefformat=parens]{subcaption}
\usepackage{bbding}
\hypersetup{
  pdftitle={Model Exposure on One Device Endangers Others: Rethinking Edge Neural Network Security across Devices},
  pdfauthor={Yuyang Zhang, Kangjie Chen, Yihui Jin, Run Wang},
  pdfsubject={Preprint}
}
\title{Model Exposure on One Device Endangers Others: Rethinking Edge Neural Network Security across Devices}
\author{
Yuyang Zhang$^{*}$ \and  Kangjie Chen$^{*}$ \and Yihui Jin$^{*}$ \and Run Wang
\affiliations
{\normalsize $^{*}$Equal contribution.}
}

\begin{document}

\maketitle
\thispagestyle{plain}

\begin{abstract}
Neural networks (NNs) are increasingly deployed on edge devices such as mobile and IoT platforms, offering low latency and improved privacy compared to cloud-based deployment.
However, on-device models embedded in mobile applications can be unpacked and exposed to adversaries, enabling \textit{cross-node attacks} in centrally distributed ecosystems, where adversarial examples crafted on one compromised device transfer effectively to other nodes sharing the same models.
To address this threat, we propose \Name, a defense framework that reduces adversarial transferability across on-device models via model obfuscation. 
The core idea of \Name is to introduce controlled model diversity while preserving original task functionality. 
Specifically, \Name applies stealthy and irreversible obfuscation at both the architectural and parameter levels through \textit{network cell obfuscation} and \textit{layer assembling}, and is applicable to diverse NN architectures, including MLPs, CNNs, and Vision Transformers.
Extensive experiments demonstrate that \Name effectively defends against cross-node attacks, achieving an average defense success of \textbf{49.19\%} and up to \textbf{81.19\%} across four adversarial attack types, with more than \textbf{33.16\%} improvement over existing defenses.
\end{abstract}

\section{Introduction}

\label{sec:intro}

In the past decades, deep learning (DL) models have been widely applied in many cutting-edge fields (\eg, image classification, speech recognition, natural language processing) and achieved impressive progress~\cite{lin2020mcunet}. These DL models are commonly deployed in cloud servers (traditional AI systems) to provide services to our users~\cite{sun2021mind,oliynyk2023know}. In the very recent few years, AI chips that are dedicated to edge devices have been developed and widely assembled in mobile and IoT devices for empowering their capability in tackling heavy computing on edge devices~\cite{ignatov2019ai}. Thus, we are witnessing massive applications with on-device models for providing many fruitful services to end users~\cite{xu2019first}. Those on-device models are gaining popularity due to their low latency in prediction and needless user data transmission, thus this paradigm becomes more efficient and safe~\cite{wang2025empowering,huang2022smart}.

\begin{figure}[t]
\centering
\includegraphics[width=\linewidth]{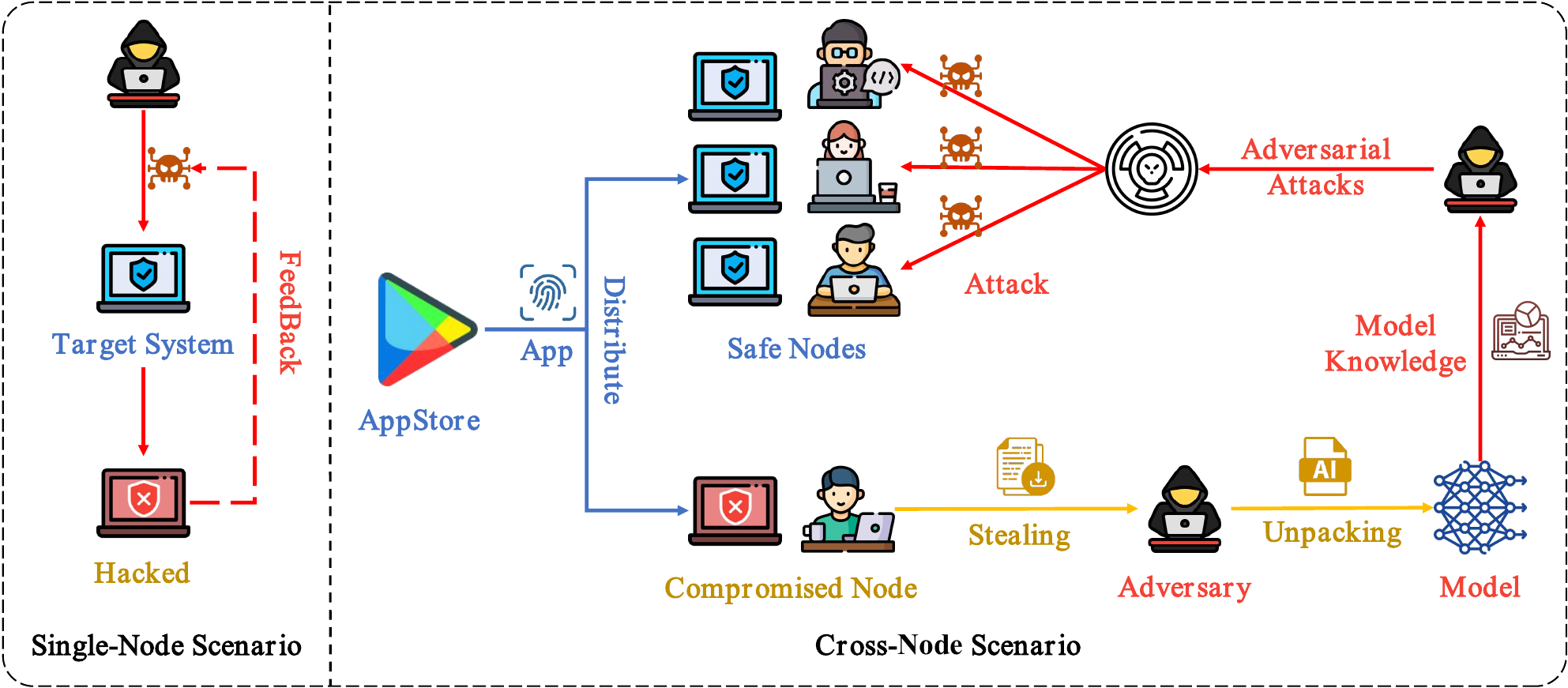}
\caption{\textbf{Threat of adversarial attacks in \textit{cross-node scenarios}.} 
Unlike single-node scenarios, which evaluate adversarial robustness solely on the targeted device and overlook threats to other identical nodes, cross-node scenarios allow an attacker to compromise one node, gain white-box access, and craft adversarial examples that successfully transfer to other devices running the same deep-learning-driven applications.
}
\label{fig:cross_node_attack}
\end{figure}

Unfortunately, numerous studies have revealed that a large quantity of edge devices are untrustable, where the applications deployed could be easily reversed by adopting the popular unpacking techniques~\cite{sun2021mind} and the embedded on-device models could be easily stolen by the popular model stealing techniques \cite{oliynyk2023know}. Thus, the on-device models are exposed to the adversary in the white-box setting which is more vulnerable to various adversarial attacks, such as evasion attacks and bit-flip attacks\cite{dong2023one}, \etc. In a more practical and real scenario, the on-device models are distributed in a centralized manner via the app store (\eg, Google Play) \cite{hu2023first} and suffer the \textit{cross-node attack}.
\textbf{Figure~\ref{fig:cross_node_attack}} shows the adversarial attacks in \textit{cross-node scenarios} against on-device models from the same mobile application. 
The attacker aims at compromising a target on-device model by unpacking the app directly or reversing the model weights to conduct a white-box adversarial attack and leverage the knowledge(\eg, model architecture, trainable weight, defense strategies) of the compromised node could be leveraged to attack the other on-device models sharing the same models or applications in black-box settings. 
For example, Huang \etal~\cite{huang2022smart} found that the adversarial examples crafted from the compromised node could be easily applied to fool the on-device models (safe nodes) from the same mobile application.
Although these target nodes are not compromised themselves, the uniformity of model deployment allows the attack to succeed in a black-box manner. This contrasts with single-node attacks, which only evaluate whether adversarial examples succeed on the targeted device, while overlooking the broader threat posed to other identical nodes. As a result, defense methods that appear effective in \textit{single-node} settings fail entirely under \textit{cross-node scenarios}.

Prior studies have shown that models with similar architectures and partially aligned parameters exhibit strong adversarial attack transferability~\cite{sun2021mind,wang2024revisiting}. 
Such transferability is a key enabler of black-box attacks. 
However, in \textit{cross-node scenarios}, the deployed models are identical in both architecture and parameters across devices. 
This complete homogeneity further amplifies the vulnerability, as adversaries can generate highly effective attacks using white-box access to any compromised node~\cite{zhou2024investigating}. 
This makes safeguarding the model’s architecture and weights from being exposed crucial for defending against the adversarial attack in \textit{cross-node scenarios}. 

A straightforward approach to mitigating transferability is to design a special model in a different architecture and retrain the model with different hyperparameters. However, re-designing and retraining the model is computationally expensive, and the illogical structure and training parameters can negatively impact the model's performance~\cite{gheisari2023deep}.
To mitigate the threat posed by model architecture leaks, existing studies introduce structural variations across different nodes. 
Existing model modification techniques~\cite{zhou2023modelobfuscator,zhou2024model} attempt to achieve this by injecting redundant components into the architecture. However, these techniques do not fundamentally disrupt the consistency of model behavior, different models still produce similar outputs for the same input, leaving them vulnerable to transferable attacks. 
Another defense strategy is to introduce perturbations into the model input~\cite{grana2020perturbing,guiga2020neural}, hidden layers~\cite{yu2021simple,yang2023simple} or perturbing the model output directly~\cite{lee2019defending,kariyappa2020defending,zhang2023apmsa}. Yet, our experiments (Section~\ref{sec:exp_results}) demonstrate that perturbation-based approaches provide only limited protection.
The lack of explicit guidelines for counteracting adversarial attacks results in a probabilistic success rate of the adversarial samples against the model. Upon repeated execution of the attack, the success rate gravitates towards 100\% rapidly.
Therefore, existing methods fail to effectively address the threat of \textit{cross-node attacks} for on-device models, necessitating a more robust defense mechanism.

In this paper, to protect the on-device models against \textit{cross-node attack}, for the first time, we propose a novel approach \Name
\footnote{\Name draws inspiration from the word \textit{veil}, referring to concealment for protection. Accordingly, \Name obscures the internal structure of models via obfuscation while preserving functionality.}
to obscure both the architecture and the weights of the models, thereby mitigating this threat.
Specifically,
to obfuscate the trainable parameters and weaken the transferability between models, we present the network cell obfuscation by injecting a set of grouped linear-correlated cells into the network without affecting the outputs of adjacent cells. 
Then the weights of the sampled cells are perturbed in a modification manner to achieve stealthy and irreversible obfuscation. 
To obfuscate the model architecture and evade the model architecture recovery, we propose the network layer obfuscation by injecting, decomposing, and fusing layers to bring the largest variations to the topology of the network and keep the performance in legitimate sample prediction. 
Our pioneer work is inspired by the widely adopted code obfuscation to defend against malicious code injection and copyright protection~\cite{sebastian2016study}.

To conduct a comprehensive evaluation of our \Name, the experiments are evaluated on several popular types of NNs with 5 powerful adversarial attacks. Experimental results demonstrate that \Name weakens the transferability significantly between obfuscated models originating from the same raw model. 
Our contributions are as follows:
\begin{itemize} %

\item We propose \Name, the first obfuscation-based defense mechanism designed to counter \textit{cross-node attacks} by significantly reducing attack transferability. \Name introduces stealthy and irreversible variations into the model’s architecture and parameters, making adversarial exploitation and model stealing attacks infeasible.

\item We develop two complementary techniques: \textit{network cell obfuscation} and \textit{network layer assembling}. The former injects grouped linear-correlated cells and perturbs their weights to obfuscate model parameters, while the latter modifies network topology via layer decomposition, injection, and fusion, ensuring architectural uniqueness and preventing model recovery.

\item We extensively evaluate \Name on three popular NNs under five adversarial attack methods and three model extraction attacks. 
Our results demonstrate that \Name achieves up to \textbf{81.19\%} and an average of \textbf{49.19\%} reduction in cross-node attack success rates, outperforming existing defenses by over \textbf{33.16\%} in most settings.

\end{itemize} 

\section{Related Work}
\label{sec:related}

\subsection{Adversarial Attack}

Adversarial attacks exploit imperceptible perturbations to manipulate the target models' prediction~\cite{liang2022adversarial}. Even minor, human-invisible modifications can lead to severe consequences, especially in security-sensitive applications such as autonomous driving~\cite{song2018physical,lovisotto2021slap}.
Depending on the attack conditions, existing attacks can be categorized into \textit{white-box} methods(\eg, FGSM, PGD, and C\&W), which achieve high-confidence misclassifications with gradient-based perturbations but assume strong adversary capabilities, and \textit{black-box} attacks, leveraging transferability or decision-boundary, such as HJSA~\cite{chen2020hopskipjumpattack} and Ghost~\cite{li2020learning}. 
For an attacker, the more knowledge they acquire about the target model, the more effective their adversarial attack becomes~\cite{liang2022adversarial}. 

Most adversarial attacks focus on \textit{single-node scenarios}, relying on the restrictive assumption that the attacker lacks access to any model parameters or architecture details beyond the immediate target node.
However, in \textit{cross-node scenarios}, adversarial attacks pose a more serious and realistic threat and are more difficult to prevent since the identical model is deployed across devices, compromising a single untrusted node grants adversaries comprehensive knowledge of the same model and its defenses on other nodes. Unlike black-box attacks, where adversaries only approximate the victim, the cross-node adversary can craft perturbations with access to internal information and defense methods of the targeted model, significantly increasing the potency.

\subsection{On-device Model Protection}

On-device models suffer various model extraction attacks and further conduct diverse efficient adversarial attacks to evade the sample prediction or data leakage~\cite{huang2022smart}. 
Recently, studies have been working on exploring effective model protection techniques to defend the aforementioned threats by data perturbation and model modification. 
Data perturbation introduces perturbations to the model layers to prevent stealing proactively. 
Gugia \etal~\cite{guiga2020neural} add perturbations to the image input using the gradient-weighted class activation mapping method. 
Model modification generally aims to protect the model structure, especially when the owner of models proposes a novel model structure that has a significant advantage in specific aspects compared with other existing models. Zhou~\etal~\cite{zhou2023modelobfuscator} proposed the ModelObfuscator, which is based on the functions and customized model structure in the TFLite, to avoid leakage of inner information of models while it still fails to evade code analysis.

However, the existing proactive defense methods are not practical for protecting the on-device models against popular adversarial attacks in \textit{cross-node scenarios}. 
The architecture and trainable parameters of the on-device models could be easily inferred via static program analysis techniques~\cite{sun2021mind,malik2024systematic}. 
Thus, the knowledge of the compromised on-device model could be leveraged for an attack to conduct large-scale and efficient adversarial attacks on other nodes in black-box settings.

\section{Motivation}

Existing defense strategies against adversarial attacks predominantly target \textit{single-node settings}, such as adversarial training~\cite{wang2024revisiting} and input-latent-output transformations~\cite{yu2021simple}, where the defense mechanisms remain hidden from the attacker, where the adversary is assumed to have no access to any internal knowledge of the model apart from the target device. These approaches offer partial robustness, particularly against adversarial examples with limited transferability.

However, in edge device systems, it is common for it to be deployed across numerous nodes for the same application, embedding an identical deep learning model. If the model on a single node is compromised, the attacker effectively gains white-box access to all nodes running the same model. 
In such cases, attackers are no longer constrained by transferability. A successful attack on one node can be directly replicated across others. Even in cases without full white-box access, adversarial examples crafted on one node can often transfer effectively to others, owing to the homogeneity of deployed models. 
Under these conditions, conventional defenses such as adversarial training lose their effectiveness. Attackers can exploit their knowledge of the model’s architecture and defense strategy to generate highly tailored perturbations that circumvent the adversarial robustness mechanisms. 
Similarly, randomized neural networks or input-output transformations become ineffective, as attackers can estimate their expectations locally and still generate successful adversarial examples.

\begin{figure}[t]
  \centering
  \begin{subfigure}[t]{0.47\columnwidth}
    \centering
    \includegraphics[width=\linewidth]{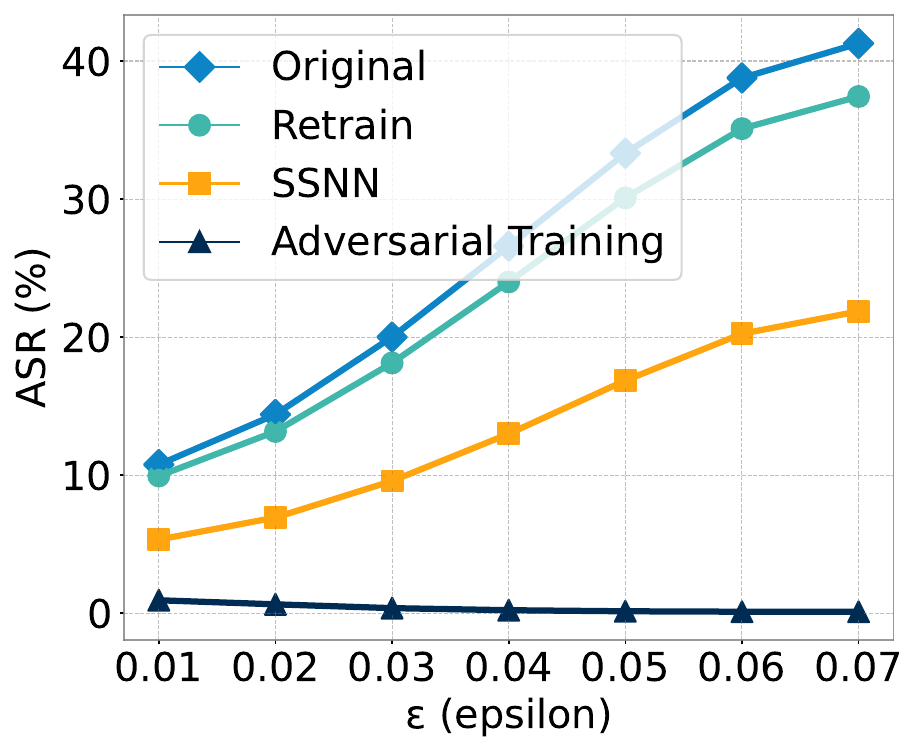}
    \caption{Single-Node}
    \label{fig:motivation1}
  \end{subfigure}
  \hfill
  \begin{subfigure}[t]{0.47\columnwidth}
    \centering
    \includegraphics[width=\linewidth]{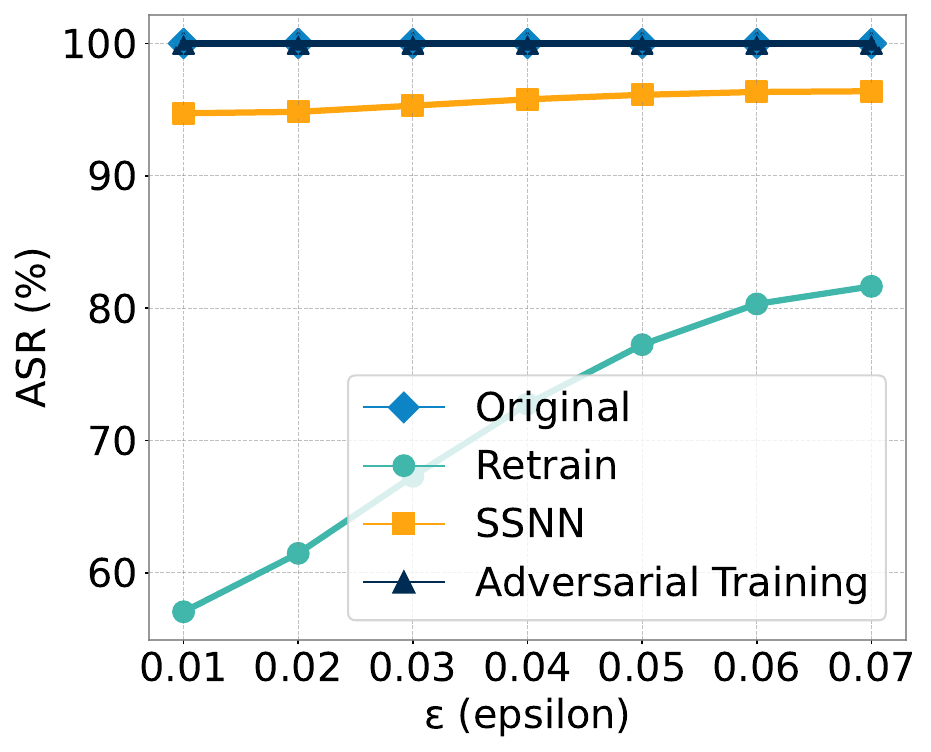}
    \caption{Cross-Node}
    \label{fig:motivation2}
  \end{subfigure}
  \caption{\textbf{Motivation studies.} The result of 3 defense strategies against transfer-based attacks in \textit{single-} and \textit{cross-node scenarios}.}
  \label{fig:motivation}
\end{figure}

We conduct preliminary experiments on the MLP model to evaluate the effectiveness of existing defenses against transfer-based attacks~\cite{TransferAttack2025Ge} in both \textit{single-node} and \textit{cross-node} scenarios.
As shown in \textbf{Figure~\ref{fig:motivation}}, $\epsilon$ denotes the attack perturbation strength, and \textit{ASR} measures the success rate of adversarial attacks.
The results in \textbf{Figure~\ref{fig:motivation}\subref{fig:motivation1}} show that existing defense methods exhibit a certain degree of robustness against adversarial attacks in the \textit{single-node settings}. However, as illustrated in \textbf{Figure~\ref{fig:motivation}\subref{fig:motivation2}}, all methods fail to provide effective protection under cross-node attack scenarios, except for retraining the model with varied hyperparameters.

In summary, most existing defenses are designed for single-node threats and overlook the security risks posed by identical deployments across nodes, leading to near-complete failure under cross-node attacks, where retraining remains the only effective but costly solution.

\section{Threat Model}

In edge systems, application distribution often involves deploying identical models across numerous user devices. If even a single node is compromised, the adversary can gain full access to the model’s internals and defense mechanisms. This shared model exposure drastically amplifies the effectiveness of adversarial attacks under the cross-node threat model. Therefore, we consider a more realistic and scalable threat setting, the cross-node scenario, illustrated in \textbf{Figure~\ref{fig:cross_node_attack}}. 
In this setting, the model owner trains a deep learning model $\mathcal{M}$ and embeds it into applications for deployment on edge devices.
Following the assumptions commonly adopted in prior works on extraction techniques~\cite{zhou2023modelobfuscator,zhou2024model} and adversarial attacks~\cite{chen2020hopskipjumpattack} against edge AI applications, we assume that adversaries can reverse engineer the application package files (\eg, APK for Android, IPA for iOS) using off-the-shelf tools such as Apktool and IDA Pro.
This provides the potential for the adversaries to extract the internal information of models via various model extracting techniques. 
For example, the model $\mathcal{M}$ could be extracted by calling the system APIs of the development format directly~\cite{zhou2024model}.
Then, the adversary $\mathcal{A}$ opens the white-box access to the target on-device models, and the mobile device is compromised further. 
Different from transfer-based adversarial attacks, the cross-node attacks occur in the white-box condition, where the attackers have full information about the target models and defenses, which makes defense more challenging.

\begin{figure*}[t]
    \centering
    \includegraphics[width=\linewidth]{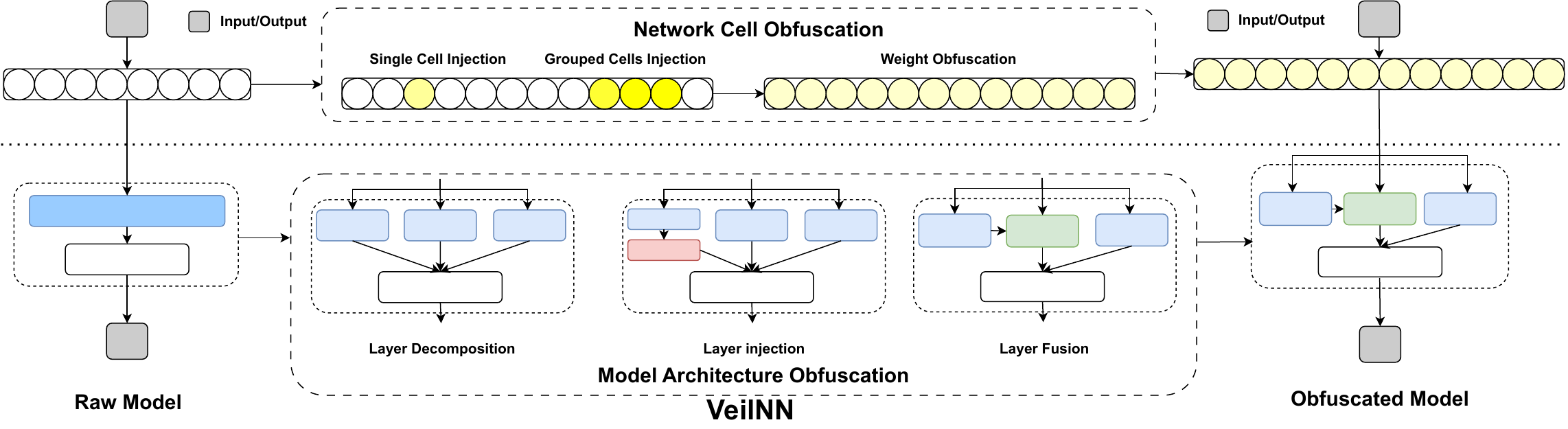}
    \caption{\textbf{The workflow of \Name.} The top panel network cell obfuscation includes single-cell injection, grouped cell injection, and weight obfuscation to introduce variations into the training parameters. The bottom panel model architecture obfuscation includes layer decomposition, layer injection, and layer fusion to resist the raw model recovery from the obfuscated model.}
    \label{fig:VeilNN}
\end{figure*}

\section{Methodology}
\label{sec:Methodology}

\subsection{Overview}

In this paper, we present a novel method by introducing model obfuscation into the target NNs to defend the \textit{cross-node attack} against on-device models by launching large-scale and efficient adversarial attacks in a black-box setting. Our basic insight lies in enlarging the differences between models to weaken the transferability, instead of employing the various model protection techniques to prevent leakage of model internals or adversarial training.

During the distribution of the applications assembled with on-device models, we apply our proposed \Name to generate a proprietary obfuscated model for each user. The obfuscated models originated from the same application have individual model architecture and trainable parameters while maintaining comparable performance.
As shown in \textbf{Figure~\ref{fig:VeilNN}}, given a target raw model, \Name first extracts its architecture and trainable parameters. It then applies the proposed model obfuscation method, leveraging \textit{network cell} and \textit{layer obfuscation} to introduce variations in the trainable parameters and obfuscate the model architecture by reshaping and rearranging layers.

\subsection{Network Cell Obfuscation}

\textbf{Cells injection} introduces injected cells in a stealthy manner, ensuring that the variations remain unidentifiable to users. \textbf{Figure~\ref{fig:VeilNN}} illustrates the injection of grouped cells into the target model.  
To prevent any impact on the outputs of adjacent layers, the parameters of the injected cells are designed to be linearly correlated, thereby preserving the functionality of the original model.
To ensure that the output of adjacent layers remains unaffected, we impose the constraint in \textbf{Equation~(\ref{eq:group_units_constraint})} on the $q$ injected neurons $[N_1, \dots, N_q]$ at layer $L^{i}$. 
\begin{equation} \label{eq:group_units_constraint}
\setlength\abovedisplayskip{3pt}%
\setlength\belowdisplayskip{3pt}
\begin{split}
Z^{{i+1}}_{1,n_{i+1}} = X^{{i}}_{1,q} W^{{i+1}}_{q,n_{i+1}}
\end{split}
\end{equation}

Moreover, to ensure the grouped cells injection does not affect adjacent layers, $Z^{{i+1}}_{1,n_{i+1}}$ must be a zero tensor for the injected grouped cells. Notably, the trainable parameters of the injected grouped cells should be designed to follow the same distribution as the original cells, thereby achieving stealthy obfuscation.

\textbf{Weight obfuscation} aims to achieve irreversible obfuscation, preventing adversaries from recovering the raw model from the obfuscated models and performing cross-node attacks. 
A sophisticated attacker cloud exploit \textbf{Equation~(\ref{eq:group_units_constraint})} to extract the injected cells from each layer, \eg, through model compression~\cite{2024microsoft}. To counter this, we propose to obfuscate the entire weight space with $\mathcal{F}$:
\begin{equation*}
\mathcal{L}' =\mathcal{F}_{\pi}(\mathcal{L},p,\alpha)
\end{equation*}
where $\mathcal{L}$ represents the target layer, $p$ denotes the parameter index within $\mathcal{L}$, $\alpha$ signifies the modification strength, and $\pi$ encapsulates the control settings, encompassing aspects such as the type of obfuscating perturbation distribution and modification amplitude limit, among other factors.
In such a scenario, the computational complexity of inferring the grouped injected neurons is \textbf{O($2^n$)} where $n$ is the number of cells in the obfuscated layer.

To further enhance model security, we introduce an irreversible obfuscation technique through weight obfuscation to minimize the risk of raw model recovery. Specifically, we apply Singular Value Decomposition (SVD)~\cite{Vlad2024Singular} to modify the parameters of target cells, thereby disrupting the decision boundary and eliminating linear dependencies among obfuscated components:
\begin{equation} \label{eq:svd}
    \mathbf{W}=\mathbf{U} \boldsymbol{\Sigma} \mathbf{V}^{T}, \quad \mathbf{\Sigma}=\operatorname{diag}\left(\sigma_{1}, \sigma_{2}, \ldots, \sigma_{r}\right).
\end{equation}
Let the two adjusted models be:
\begin{equation*}
\mathbf{W}_1 = \mathbf{U}(\mathbf{\Sigma} + \mathbf{\Delta}_1)\mathbf{V}^T, \quad 
\mathbf{W}_2 = \mathbf{U}(\mathbf{\Sigma} + \mathbf{\Delta}_2)\mathbf{V}^T,
\end{equation*}
where $\mathbf{\Delta}_1 = \mathrm{diag}(\vec{\delta}^{(1)})$, $\mathbf{\Delta}_2 = \mathrm{diag}(\vec{\delta}^{(2)})$ are independent noise matrices, with entries satisfying:
\begin{equation*}
|\delta_i^{(k)}| < \varepsilon \sigma_i \quad (k=1,2; \ \varepsilon \ll 1).
\end{equation*}
For any input $\mathbf{x}$, the adversarial perturbations $\vec{\eta}_1, \vec{\eta}_2$ are derived as:
\begin{equation*}
\vec{\eta}_k = \arg\max_{\|\vec{\eta}\| \leq \varepsilon} \left\| \nabla_{\mathbf{x}} \mathbf{W}_k(\mathbf{x}) \vec{\eta} \right\|_2 \quad (k=1,2).
\end{equation*}
The adversarial direction angle $\theta$ is defined as:
$$
\cos\theta = \frac{\langle \vec{\eta}_1, \vec{\eta}_2 \rangle}{\|\vec{\eta}_1\|_2 \|\vec{\eta}_2\|_2} = \frac{\mathrm{tr}(\mathbf{\Sigma}_1 \mathbf{\Sigma}_2)}{\sqrt{\mathrm{tr}(\mathbf{\Sigma}_1^2) \mathrm{tr}(\mathbf{\Sigma}_2^2)}},
$$
where $\mathbf{\Sigma}_1 = \mathbf{\Sigma} + \mathbf{\Delta}_1$, $\mathbf{\Sigma}_2 = \mathbf{\Sigma} + \mathbf{\Delta}_2$.
For an adversarial sample $\mathbf{x}_{\text{adv}} = \mathbf{x} + \vec{\eta}_1$ generated from $\mathbf{W}_1$, its attack success rate on $\mathbf{W}_2$ is bounded by $P_{\text{success}} \leq \kappa \cdot \cos\theta$, where $\kappa$ quantifies the task-dependent relevance of adversarial direction alignment.

\subsection{Network Layer Assembling}
An attacker can reconstruct the original cells and weights by statistically analyzing models across multiple devices.
Our experimental results (Section Main Results) reveal that an attacker with access to the original model weights can launch powerful cross-node attacks.
Consequently, obfuscating only the cells and their weights is insufficient to mitigate this threat. 
To further enhance security, it is crucial to disrupt correlations between individual model weights, which can be achieved through layer modification.
Therefore, we propose \textit{network layer assembling} to introduce structural variations into the network. Specifically, as illustrated in \textbf{Figure~\ref{fig:VeilNN}} we implement this obfuscation strategy through \textit{layer injecting}, \textit{layer decomposing}, and \textit{layer fusion}, ensuring that these modifications do not degrade performance on legitimate sample predictions.  %

A layer consists of a collection of cells and plays a fundamental role in executing specific functions. For example, Conv2D layers extract spatially local feature maps from image data. By obfuscating layers, we aim to confuse adversaries attempting to analyze and interpret the functional roles of different network layers.
Therefore, we explore three key techniques: (1) \textit{layer injection} introduces additional inference processes and perturbs selected cells. (2) \textit{layer decomposition} introduces new computational paths within the network topology. (3) \textit{layers fusion} removes certain computational paths to alter network structure. Next, we elaborate on each of them in detail.

\textbf{Layer injection} aims to introduce additional computational processes to obfuscate predictions or intermediate data. The injected layers further complicate the computational graph of the target model and prevent knowledge leakage from the obfuscated model.
To enhance stealthiness, instead of injecting standalone layers—which could be easily detected—we introduce layers in pairs, allowing greater flexibility in weight design while maintaining an inconspicuous structure.
We define $\oplus$ as a summation operation, and the constraint for injected layers is formulated as:
\begin{equation*} \label{eq:Additional_layers_injection}
\setlength\abovedisplayskip{3pt}%
\setlength\belowdisplayskip{3pt}
\begin{split}
X^{{i-1}}_{1,n_{i-1}}\Leftrightarrow  
(X^{{i-1}}_{1,n_{i-1}}W_{1} + B_{1})  \oplus \cdots  \oplus
(X^{{i-1}}_{1,n_{i-1}}W_{n} + B_{n})
\end{split}
\end{equation*}
To preserve the performance in legitimate sample prediction, the output of the injected layers must be equivalent to the original input $X^{{i-1}}_{1,n_{i-1}}$.

\textbf{Layer decomposition} aims to introduce an additional computational branch into the network topology, increasing its complexity and making it more difficult for an adversary to extract the original model architecture from the obfuscated version. 
Suppose that $layer_{i}$ is decomposed into two layers containing $k$ and $n_{i}-k$ neurons, respectively. We define $ \Join $ as a concatenate operator, and the constraint for layer decomposition is formulated as follows:
\begin{equation} \label{eq:layer_decompositions_constraint}
\setlength\abovedisplayskip{3pt}%
\setlength\belowdisplayskip{3pt}
\begin{split}
X^{{i-1}}_{1,n_{i-1}}W^{{i}}_{m_{i},n_{i}} &\Leftrightarrow  
(X^{{i-1}}_{1,n_{i-1}}W^{{i}}_{m_{i},k} + B^{{i}}_{1,k})  \\ 
&\Join (X^{{i-1}}_{1,n_{i-1}}W^{{i}}_{m_{i},n_{i}-k} + B^{{i}}_{1,n_{i}-k})
\end{split}
\end{equation}
where $m_{i}$ equals $n_{i-1}$. The \textbf{Equation~(\ref{eq:layer_decompositions_constraint})} indicates that the output of the decomposed layers is equivalent to the raw layer.

\textbf{Layer fusion} is the inverse process of layer decomposition. It enhances model obfuscation by merging two unrelated layers, integrating their independent computational processes, and further complicating both the architecture and execution flow of the model.
In layer fusion, the input to the fused layers can either be the same or different, leading to two possible fusion approaches. For layers sharing the same input $X$, their weights can be merged directly as follows:
\begin{equation*} 
\setlength\abovedisplayskip{3pt}%
\setlength\belowdisplayskip{3pt}
\begin{split}
X_{A} W_{A}\Join X_{B} W_{B}
\Leftrightarrow 
\begin{bmatrix}
X_{A} & X_{B}
\end{bmatrix}
\begin{bmatrix}
W_{A}&\\
&W_{B}
\end{bmatrix}
\end{split}
\end{equation*}
To ensure that layer fusion does not affect the inputs of adjacent layers, the composite layer must merge the outputs of the raw layers.

\begin{figure*}[t]
    \centering
    \begin{subfigure}{0.21\textwidth}
        \centering
         \includegraphics[height=3.09cm]{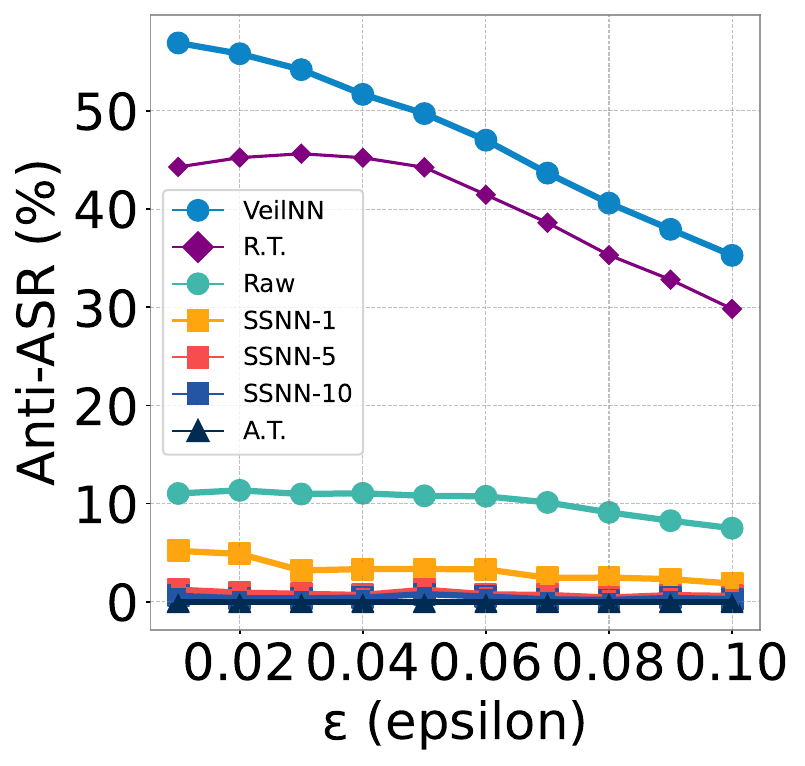}
        \caption{\textbf{Defend Methods}}
        \label{fig:epsilon}
    \end{subfigure}\hfill
    \begin{subfigure}{0.36\textwidth}
        \centering
        \includegraphics[height=3.09cm]{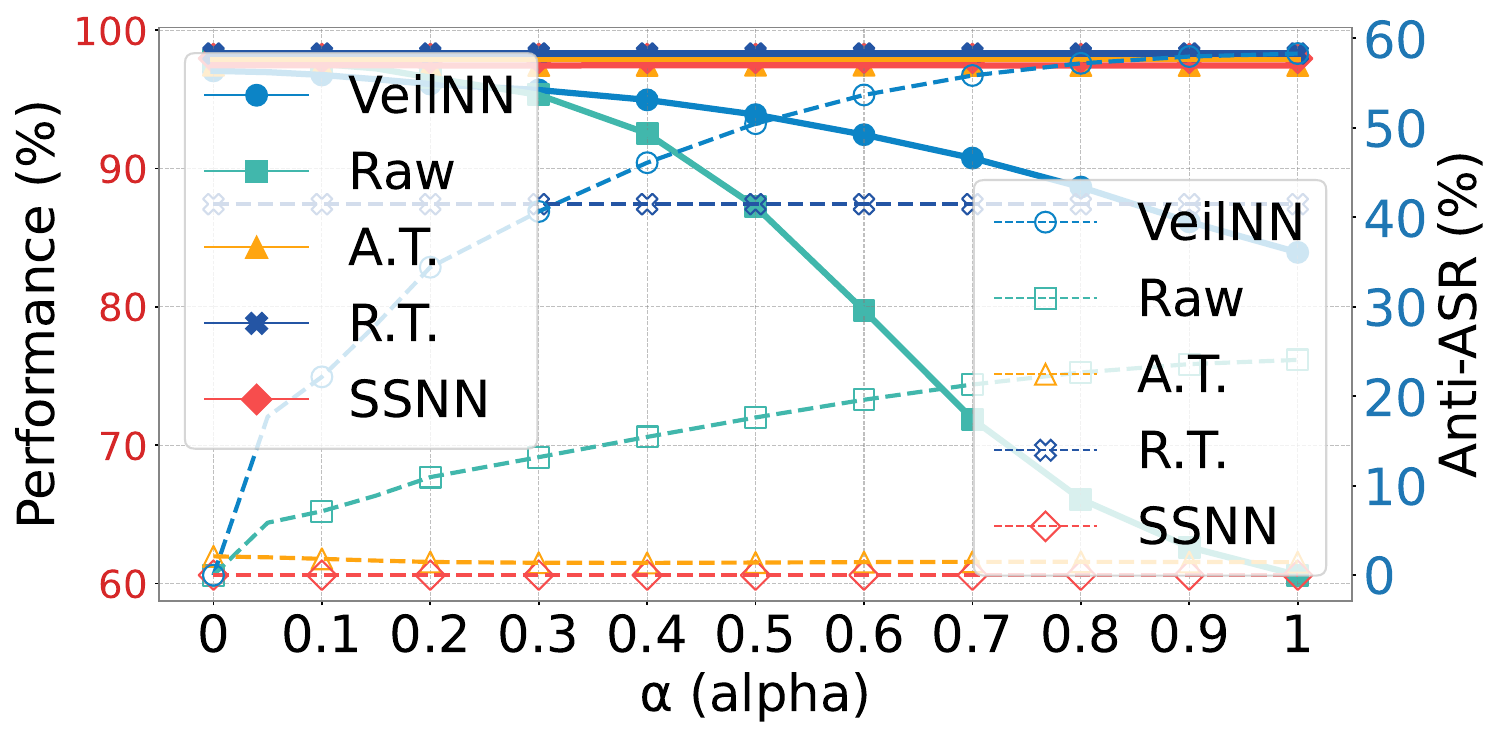}
        \caption{\textbf{Performance vs Anti-ASR}}
        \label{fig:performance}
    \end{subfigure}\hfill
    \begin{subfigure}{0.21\textwidth}
        \centering
        \includegraphics[height=3.09cm]{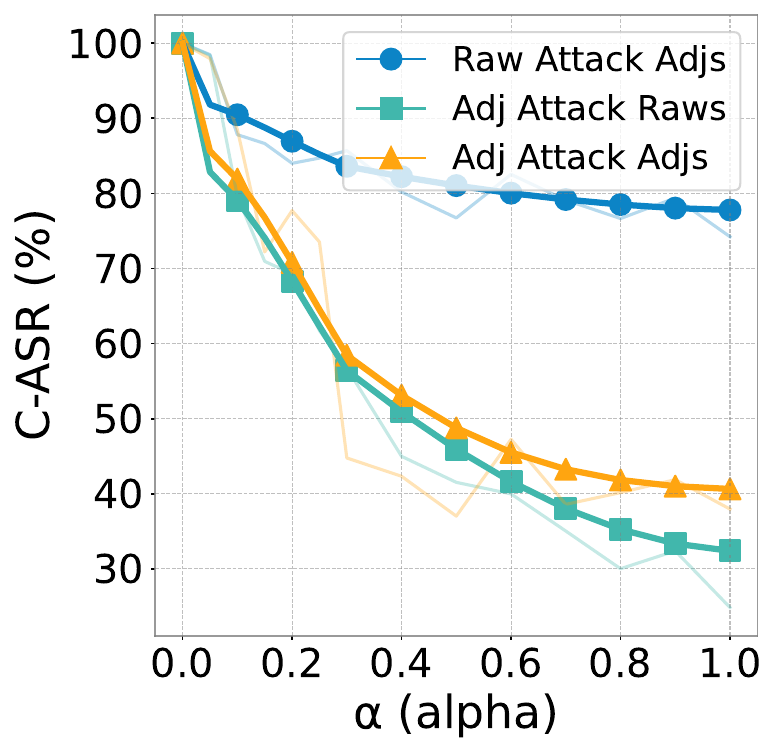}
        \caption{\textbf{Knowledge}}
        \label{fig:knowledge}
    \end{subfigure}\hfill
    \begin{subfigure}{0.21\textwidth}
        \centering
        \includegraphics[height=3.09cm]{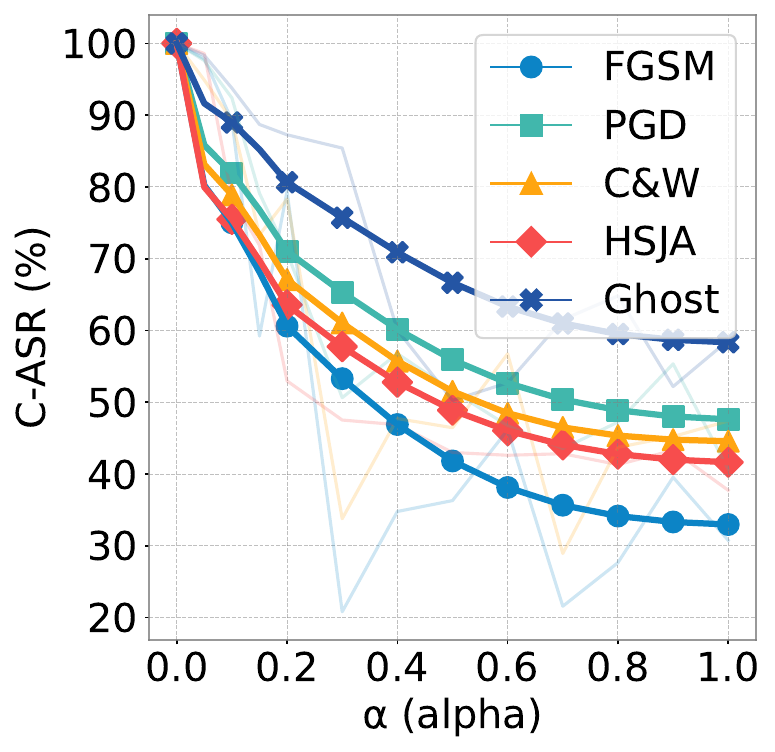}
        \caption{\textbf{Attack Methods}}
        \label{fig:attack_method}
    \end{subfigure}
    \caption{\textbf{Main results.} 
    Figure~\subref{fig:epsilon}: We compare the Anti-ASR of \Name with baselines under adversarial attacks of ten different intensities, characterized by $\epsilon$.
    Figure~\subref{fig:performance}: We compare the performance on original tasks and their defense capability of the model obfuscated by \Name under various $\alpha$.
    Figure~\subref{fig:knowledge}: We analyze the C-ASR of cross-node attacks under three different levels of adversarial knowledge.
    Figure~\subref{fig:attack_method}: We examine \Name's defense effectiveness against cross-node attacks employing five distinct adversarial algorithms.}
    \label{fig:main}
\end{figure*}

\section{Experiments}
\label{sec:exp_results}

\subsection{Experiment Setup}\label{sec:setup}
Given the complexity of visual data, models deployed on edge devices are especially vulnerable to adversarial attacks, making them a frequent target of increasingly sophisticated threat techniques~\cite{li2024adversarial}. Accordingly, we center our evaluation on visual models and the associated attack and defense mechanisms. We structure our experiment around three key aspects: Main Results, Anti Model Extracting, and Ablation Studies. The ablation results are reported in the Supplementary Material.

\noindent\textbf{Datasets and models.}
Our experiments are conducted on two classic datasets, \ie, MNIST~\cite{deng2012mnist}, CIFAR~\cite{krizhevsky2012imagenet} and Imagenet~\cite{deng2009imagenet}, using three popular neural networks, \ie, MLP~\cite{nie2021mlp}, AlexNet~\cite{krizhevsky2012imagenet}, ResNet~\cite{he2016deep} and ViT~\cite{dosovitskiy2020image}.

\noindent\textbf{Adversarial attacks for evaluation.}
For white-box attacks, we evaluate our method against the one-step FGSM attack~\cite{goodfellow2014explaining}, the multi-step PGD attack~\cite{madry2017towards}, and the optimization-based C\&W attack~\cite{carlini2017towards}. For black-box attacks, we employ the query-based HSJA attack~\cite{chen2020hopskipjumpattack} and the transfer-based attack, Ghost attack~\cite{li2020learning}. 
Considering the security of Edge-AI models, we implement untargeted attacks to emphasize the importance of model accuracy in real-world scenarios.

\noindent\textbf{Baselines.} 
For anti-cross-node attacks, we compare our method against retraining model (R.T.), stochastic neural networks (SSNN)~\cite{yu2021simple}, adversarial training (A.T.)~\cite{madry2017towards}, and raw weight modification~\cite{szentannai2020preventing}. 
For model extraction defenses, we use ModelObfuscation~\cite{zhou2023modelobfuscator} and CustomDLCoder~\cite{zhou2024model} as baselines.

\noindent\textbf{Evaluation metric.}
\ding{172} Cross-node attack success rate (C-ASR): To evaluate attack performance, we compute C-ASR by generating adversarial samples on each model and measuring the average attack success rate on other edge models.  
\ding{173} Anti-cross-node attack success rate (Anti-ASR): We assess defense effectiveness using Anti-ASR, defined as $1 - \textit{C-ASR}$.  
\ding{174} Model structure similarity: We quantify the structural similarity between models using the Neighborhood Subgraph Pairwise Distance Kernel (NSPDK)~\cite{costa2010fast} from GraKeL~\cite{GraKeL}.

\subsection{Main Results}\label{sec:main_results}

In this section, we evaluate the defensive effectiveness of \Name against cross-node attacks. 
Specifically, we assess \ding{172} its performance in comparison with state-of-the-art (SOTA) defense methods under standardized evaluation protocols, and \ding{173} its robustness under different attack conditions, including varying density of adversarial knowledge and attack algorithms. \textbf{Figure~\ref{fig:main}} provides an overview of the main results.

\noindent\textbf{Comparison with existing cross-node defense methods.}
We first compare our proposed method against three popular baselines in the cross-node scenario, including model retraining, stochastic architecture~\cite{yang2023simple}, and the raw weight modification method~\cite{szentannai2020preventing}. 
Notably, SSNN~\cite{yang2023simple} introduces stochastic layers, meaning the same adversarial sample can yield different effects across multiple trials. Additionally, we evaluate adversarial training, which enhances robustness by incorporating adversarial examples during training. 
To ensure a fair comparison, we focus on single-instance attacks.

As shown in \textbf{Figure~\ref{fig:main}\subref{fig:epsilon}}, both \Name and Retrain demonstrate strong resistance to adversarial attacks, with success rates dropping from 56.92\% to 35.29\% and 44.29\% to 29.83\%, respectively, as $\epsilon$ increases. It is noteworthy that our approach perturbs the decision boundaries similarly to Low-Rank Adaptation techniques~\cite{liu2024dora}, but utilizes random datasets. This results in a more effective defense compared to retraining on the original dataset, which often produces decision boundaries similar to those of the original model.
In contrast, Raw, Adversarial Training, and SSNN perform poorly. Adversarial training consistently records a 0\% success rate, while SSNN achieves at most 5.19\%, dropping to 1.23\% and 0.78\% under repeated attack attempts (5 and 10 trials). In summary, while retraining-based defenses tend to preserve decision boundaries close to those of the original model, \Name introduces diversified boundary perturbations that remain effective under increasing attack strength, leading to consistently stronger resistance across all evaluated settings.

\noindent\textbf{Effect of obfuscation intensity on model performance.}
Further, we evaluate the impact of our approach on the model's image recognition performance. 
We assessed the effect of the intensity parameter $\alpha$ on performance, using randomized modification of raw weights as a baseline. The results are presented in \textbf{Figure~\ref{fig:main}\subref{fig:performance}}. 
 The experimental data demonstrate that our method effectively preserves the model's performance when compared to introducing random parameters into the weights, 
ensuring that the loss of predictions remains below 5\% accuracy when $\alpha \leq 0.6$, where the loss of raw weight modification is below 22.5\%. 
However, compared to SSNN and Adversarial Training, our method introduces a performance loss that remains below 2.5\% accuracy when $\alpha < 0.6$. Whereas method SSNN has its performance loss at 0.10\%, Adversarial Training will not have any loss.
The deviation of the obfuscated models on the original task remains minimal, averaging only 3.29\% compared to the raw model, where $\alpha \in [0.4, 0.6]$.
Overall, these results indicate that \Name achieves a favorable balance between defensive effectiveness and model stability, introducing only limited performance degradation while maintaining robust protection against cross-node attacks.

\noindent\textbf{Effect of adversary’s knowledge on cross-node attack.} In edge systems, cross-node attack effectiveness varies based on the attacker’s knowledge. We evaluate three adversarial scenarios: \ding{172} The attacker knows the model deployed on their own edge device and uses it to compromise other devices. \ding{173} The attacker has access to the raw model and directly targets edge devices. \ding{174} The attacker has knowledge of a specific edge device, allowing them to attack the raw model. 
As shown in \textbf{Figure~\ref{fig:main}\subref{fig:knowledge}}, adversarial samples are most transferable when the attacker possesses raw model knowledge, with C-ASR of 80.03\%.
In contrast, if the attacker only has access to a specific node, their ASR drops significantly to 45.51\% for other nodes and 41.61\% against the raw model (for $\alpha = 0.6$).
These results underscore the importance of obfuscating architecture and weights to disrupt adversarial transferability and mitigate cross-node attacks.

\noindent\textbf{Effectiveness against different attack algorithms.}
Adversarial attack algorithms can be broadly categorized into two types based on their access conditions: gradient-based attacks and transfer-based attacks.
For gradient-based attacks, we employ three widely used white-box attacks, \ie, FGSM, PGD, and C\&W and one black-box attacks HSJA, as the attack algorithms.
For \Name, a natural adaptive attack is the transfer-based attack, in which adversaries leverage extracted knowledge to train surrogate models that are similar to the target model, and use them to generate adversarial examples that compromise the target nodes. Accordingly, we further evaluate \Name against the transfer-based method, Ghost.

As shown in \textbf{Figure~\ref{fig:main}\subref{fig:attack_method}}, \Name exhibits strong defense against both gradient-based and transfer-based attacks. As the obfuscation intensity $\alpha$ increases, indicating a higher degree of distortion of the decision boundary, more adversarial samples become ineffective, causing the C-ASR of the cross-node attack to decrease gradually.
In the optimal obfuscation range ($\alpha \in [0.4, 0.6]$), the overall Anti-ASR of \Name on five attacks was 49.19\%, where 57.71\% for FGSM, 43.74\% for PGD, 48.09\% for C\&W, 50.75\% for HSJA and 45.65\% for Ghost. 
Overall, these results indicate that \Name consistently mitigates cross-node attacks across diverse attack algorithms, demonstrating robust and algorithm-agnostic defensive effectiveness within a practical obfuscation range.

\subsection{Anti Model Extracting}

\begin{figure}[t]
    \centering
    \includegraphics[width=\linewidth]{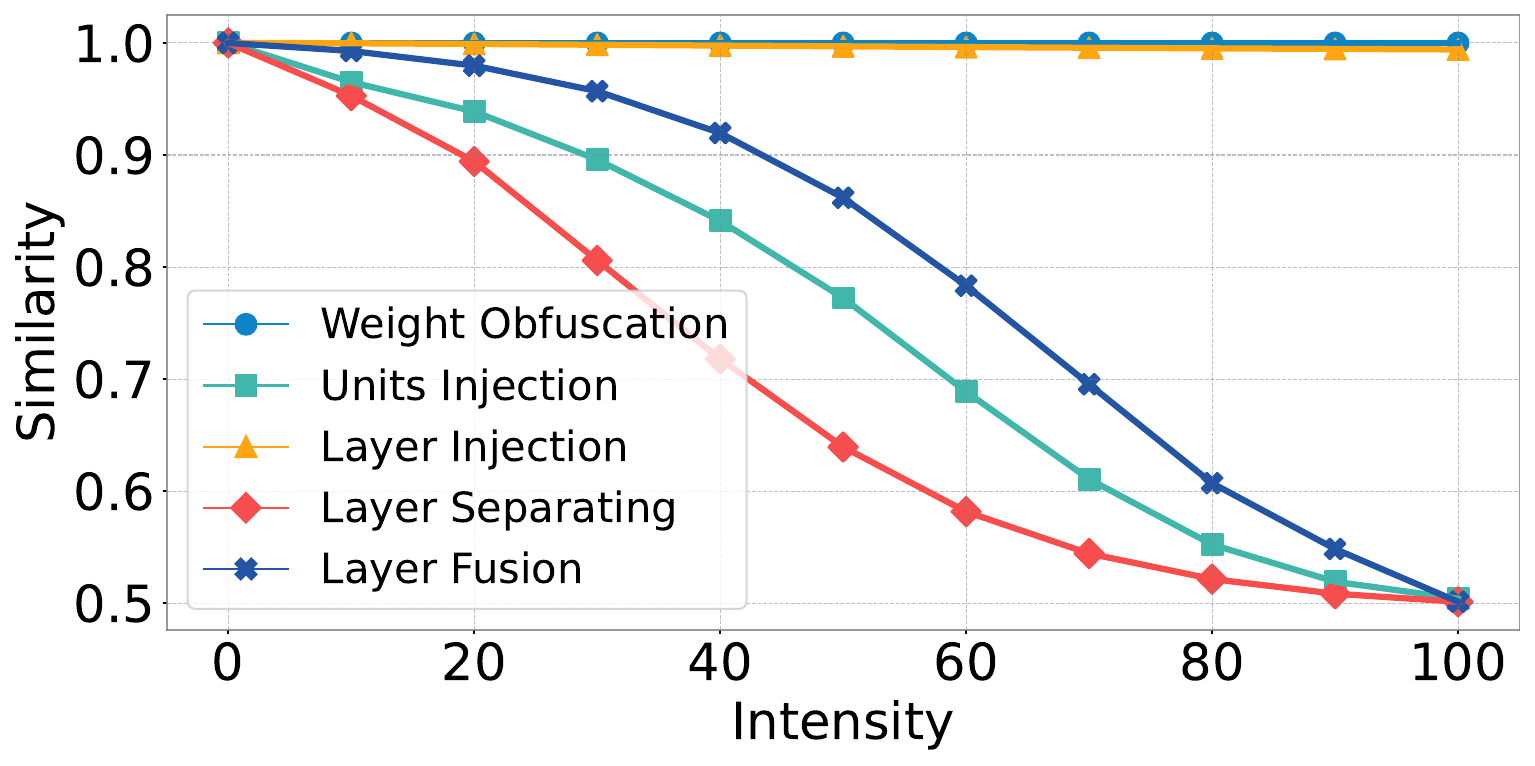}
    \caption{\textbf{Impact of different model obfuscation strategies on model structural similarity.}}
    \label{fig:model_similarity}
    \vspace{-10pt}
\end{figure}

\noindent\textbf{Effectiveness of model obfuscation.}
To explore which model obfuscation strategy introduces the greatest variations in model architecture, we conduct experiments on four distinct neural networks: MLP, Alexnet, ResNet, and Visual Transformer (ViT). We measure the structural similarity between the raw and obfuscated models using the Neighborhood Subgraph Pairwise Distance Kernel (NSPDK)~\cite{costa2010fast} in GraKeL~\cite{GraKeL}.
As illustrated in \textbf{Figure~\ref{fig:model_similarity}}, different obfuscation strategies have varying effects on model architecture: (1) Weight obfuscation and layer injection introduce the least structural variation. (2) Layer decomposition has the most significant impact, substantially altering the model’s computational graph. 
(3) Network cell injection provides a moderate level of obfuscation, disrupting model structure while maintaining stability.

\noindent\textbf{Resistance against model extraction attacks.}
As shown in \textbf{Figure~\ref{fig:main}\subref{fig:knowledge}}, an attacker with access to the raw model information can more effectively compromise the model on the edge device. Therefore, we further evaluated the threat posed by an attacker utilizing multiple raw model structures and permission weighting on the edge device. Specifically, we follow the experimental settings from CustomDLCoder~\cite{zhou2024model} and ModelObfuscation~\cite{zhou2023modelobfuscator} to evaluate extraction success rates under conversion, reverse, feature, and code analysis settings.
As summarized in \textbf{Table~\ref{tab:Anti-Extraction}}, ModelObfuscation and CustomDLCoder partially protect models but fail to prevent cross-node model leakage. \Name completely prevents direct extraction attacks and significantly reduces the success rate of code-based reconstruction. 

\begin{table}[t]  %
  \centering

    \setlength{\tabcolsep}{1mm}
    \begin{tabular}{l|ccc|c|c}
    \toprule
    \multicolumn{1}{c|}{\multirow{3}[0]{*}{Attack}} & \multicolumn{3}{c|}{Extraction} & Code & \multirow{3}[0]{*}{Cross-node} \\
    \cmidrule(lr){2-4} 
          & C.V. & R. & F. & Analyzing  &  \\
          
    \hline
    TVM   & \Checkmark & \XSolidBrush & \Checkmark\kern-1.2ex\raisebox{1ex}{\rotatebox[origin=c]{125}{\textbf{--}}}  & \Checkmark & \XSolidBrush \\
    Obfuscator & \Checkmark & \Checkmark & \Checkmark & \Checkmark\kern-1.2ex\raisebox{1ex}{\rotatebox[origin=c]{125}{\textbf{--}}}  & \XSolidBrush \\
    CustomDLCoder & - & \Checkmark & \Checkmark & \Checkmark & \XSolidBrush \\
    \textbf{Ours}  & \Checkmark & \Checkmark & \Checkmark & \Checkmark & \Checkmark \\
    \bottomrule
    \end{tabular}%
    \caption{\textbf{Defend against extracting raw model.} 
    \Checkmark: This extraction method is unable to pilfer information in accordance with the defense. `-': This method is not applicable. \Checkmark\kern-1.2ex\raisebox{1ex}{\rotatebox[origin=c]{125}{\textbf{--}}}: Partial information exposed. \XSolidBrush: Defense failure. We use C.V., R., and F. to denote Conversion, Reverse, and Feature extraction, respectively.}
  \label{tab:Anti-Extraction}%
\end{table}%

\section{Conclusion}

In this paper, we identify the cross-node attack as a critical security threat to on-device neural networks, where adversaries leverage a compromised node to launch large-scale adversarial attacks. To mitigate this threat, we propose \Name, a novel model obfuscation framework that diversifies deployed models by modifying both architecture and trainable parameters, thereby reducing attack transferability.
Extensive experiments on three neural network architectures (including MLP, ResNet, ViT) demonstrate that \Name effectively mitigates cross-node attacks, reducing the attack success rate by \textbf{49.19\%} on average and up to 81.19\% while preserving model performance.
Our findings highlight the potential of model obfuscation as a practical and scalable defense mechanism for securing on-device models.

\bibliographystyle{named}
\bibliography{ijcai26}

\end{document}